\documentclass[aps,preprint]{revtex4}
\usepackage{graphicx}
\usepackage{amssymb}
\usepackage{amsmath}
\usepackage{color}
\usepackage{enumerate}
\usepackage{bm}
\begin{document}

\title{Magnetic Relaxation in Two Dimensional Assembly of Dipolar Interacting Nanoparticles}
\author{Manish Anand}
\email{itsanand121@gmail.com}
\affiliation{Department of Physics, Bihar National College, Patna University, Patna-800004, India.}

\date{\today}

\begin{abstract}
Using the two-level approximation of the energy barrier, we perform extensive kinetic Monte Carlo simulations to probe the relaxation characteristics in a two-dimensional ($L^{}_x\times L^{}_y$) array of magnetic nanoparticle as a function of dipolar interaction strength $h^{}_d$, aspect ratio $A^{}_r=L^{}_y/L^{}_x$, and temperature $T$. In the case of weak dipolar interaction ($h^{}_d\approx0$) and substantial temperature, the magnetic relaxation follows the N\'eel Brown model as expected. Interestingly, the dipolar interaction of enough strength is found to induce antiferromagnetic coupling in the square arrangement of MNPs ($A^{}_r=1.0$), resulting in the fastening of magnetic relaxation with $h^{}_d$. There is also a rapid increase in relaxation even with $A^{}_r<100$ above a particular dipolar interaction strength $h^{\star}_d$, which gets enhanced with $A^{}_r$. Remarkably, there is a slowing down of magnetic relaxation with $h^{}_d$ for the highly anisotropic system such as linear chain of MNPs.  It is because the dipolar interaction induces ferromagnetic interaction in such a case. The thermal fluctuations also affect the relaxation properties drastically. In the case of weak dipolar limit, magnetization relaxes rapidly with $T$ because of enhancement in thermal fluctuations. The effect of dipolar interaction and aspect ratio on the magnetic relaxation is also clearly indicated in the variation of N\'eel relaxation time $\tau^{}_N$. In the presence of strong dipolar interaction ($h^{}_d>0.3$) and $A^{}_r=1.0$, $\tau^{}_N$ decreases with $h^{}_d$ for a given temperature. On the other hand, there is an increase in $\tau^{}_N$ with $h^{}_d$ for huge $A^{}_r$ $(>100)$. We  believe that the concepts presented in this work are beneficial for the
efficient use of self-assembled MNPs array in data storage and other related applications.

\end{abstract}

\maketitle

\section{Introduction}
Magnetic nanoparticles (MNPs) have received significant attention in recent years due to their unique magnetic properties and diverse technological applications~\cite{zhang2019,akbarzadeh2012,wang2019,tran2010,leo2018,felix2019,wu2019,kechrakos2002}. For instance, MNPs have various applications such as magnetic hyperthermia for cancer treatment, targetted drug delivery, biosensors, data storage devices, etc.~\cite{kumar2018,anand2016,yu2016,malekzad2017,gu2016,reiss2005}. In these applications, the dynamics of nanoparticles is primarily characterized by N\'eel relaxation time, which depends on various vital parameters such as particle size, anisotropy constant, temperature and magnetic interaction, etc.~\cite{talantsev2018,ilg2020,carrey2011,iacob2016}. Therefore, the study of magnetic relaxation in such a system represents a topic of great interest.

The relaxation characteristics of non-interacting MNPs are well understood as they are successfully explained using N\'eel and Brown theory of superparamagnetism~\cite{wernsdorfer1997,bedanta2013}. However, MNPs are found to interact because of dipolar interaction. Therefore, dipolar interaction affects the magnetic relaxation properties even for a low concentration of nanoparticles~\cite{garcia2000}. The dipolar interaction is long-ranged and anisotropic. Consequently, it can induce ferromagnetic or antiferromagnetic coupling between the magnetic moments depending on its relative position. The dipolar interaction plays an essential role in determining various thermodynamic and magnetic properties of crucial importance. For example, it induces spin-glass like behaviour in randomly distributed MNPs~\cite{majlis2018}. A highly anisotropic system such as a linear chain favours the head to the tail arrangement of magnetic moments~\cite{cheng2009}. In these contexts, several works have taken care of the role of dipolar interaction~\cite{ovejero2016,shen2001,tavares1999,chantrell1982,anand2018}. The dipolar interaction also dictates the ground state morphology of the assembly of MNPs. Luttinger {\it et al.} found the minimum energy state of a collection of classical dipoles to be ferromagnetic in a face-centred cubic lattice. In contrast, it is antiferromagnetic in a simple cubic lattice~\cite{luttinger1946}. In two dimensions, a square lattice of classical point dipoles exhibits an antiferromagnetic arrangement of the moments, while for a triangular lattice, the arrangement is ferromagnetic~\cite{macisaac1996,politi2006}.
The relaxation properties of interacting MNPs depend not only on the strength of dipolar interaction but also on the detailed MNPs arrangement and anisotropy axes orientations~\cite{dejardin2011}.
For many decades, relaxation phenomena have been explored because of their vital role in various technological applications and rich physics. 
However, a complete understanding is still elusive due to the frustrations induced by the dipolar interactions and the randomly oriented anisotropy axes~\cite{iglesias2002,labarta1993,balcells1997}.

Numerous works strengthen the fact that dipolar interaction strongly affects
magnetic relaxation properties~\cite{vernay2014,morup2010,dormann1988,jonsson2001,figueiredo2007,denisov2003,chamberlin2002,denisov2002,shtrikman1981,kalmykov2004,osaci2015,jonsson2004,berkov2001}. For example, Figueiredo {\it et al.} studied the magnetic relaxation using Monte Carlo simulation in the triangular assembly of MNPs~\cite{figueiredo2007}. They observed an exponential decay of magnetization for weakly interacting MNPs. Denisov {\it et al.} analyzed the relaxation properties in a two-dimensional assembly of MNPs with perpendicular anisotropy axes~\cite{denisov2003}. The dipolar interaction is found to slow down magnetic relaxation. Chamberlin {\it et al.} studied the relaxation in the dilute assembly of MNPs~\cite{chamberlin2002}. They observed non-exponential decay of magnetization even for weakly interacting MNPs. In another study, Denisov {\it et al.} probed the magnetic relaxation in two-dimensional assembly using mean-field approximations~\cite{denisov2002}. It has been shown that the magnetic relaxation in these ensembles is characterized by two different relaxation times. Using mean-field approximations, Shtrikmann and Wohlfarth studied the dependence of relaxation time on the dipolar interaction~\cite{shtrikman1981}. They observed an increase in relaxation time with an increase in dipolar interaction strength.  
Using Fokker-Planck formalism, Yuri P. Kalmykov derived an expression for
relaxation time of uniaxial superparamagnetic nanoparticles in the presence of an external magnetic field applied at an arbitrary angle with respect to the anisotropy axis of the particle, applicable for a wide range of damping parameter~\cite{kalmykov2004}. By taking into account the effect of damping, Osaci {\it et al.} derived an expression for $\tau^{}_N$ for dipolar interacting MNPs~\cite{osaci2015}. J$\ddot{\mathrm {o}}$nsson {\it et al.} also computed $\tau^{}_N$ in an assembly of dipolar interacting MNPs for a wide range of damping parameter~\cite{jonsson2004}.
Using micromagnetic Langevin simulations, Berkov and Gorn observed that uniaxial spins coupled via dipolar interaction exhibit damping effects such as variation in blocking temperature with interaction strength~\cite{berkov2001}.

In the context of a highly anisotropic system such as a linear chain of MNPs, similar observations have been made~\cite{laslett2015,laslett2016,iglesias2004,anand2019}. 
For instance, Iglesias {\it et al.} found that the relaxation behaviour changes from quasi logarithmic to power-law, increasing dipolar interaction strength~\cite{iglesias2004}. We have also shown that relaxation time crucially depends on the dipolar interaction strength and anisotropy axes orientation~\cite{anand2019}. It is evident from above that the effect of dipolar interaction on magnetic relaxation is not completely understood despite numerous work. An accurate evaluation of relaxation time is also equally important as it plays a vital role in various technological applications~\cite{lenin2019,hergt2009,kuncser2019,anand2021,rizzo1999}. For example,
Kuncser {\it et al.} observed that the amount of heat dissipation by the MNPs depends strongly on the relaxation time of the system~\cite{kuncser2019}. In a recent study, we have also shown that relaxation time is dictated by the dipolar interaction in a linear arrangement of MNPs, one of the essential quantifiers in determining the heat dissipation in such a system~\cite{anand2021}. 
Rizzo {\it et al.} also found that storage capacity is primarily decided by the relaxation time~\cite{rizzo1999}. 
Thus motivated, we report the effect of dipolar interaction and temperature on the magnetic relaxation in a two-dimensional assembly of MNPs with randomly oriented anisotropy axes in the present work. In particular, we perform extensive kinetic Monte Carlo simulation (kMC) to probe the relaxation characteristics as a function of dipolar interaction strength, aspect ratio and temperature in a two-dimensional array of nanoparticles.

The kMC simulation is now widely used to study the time-dependent properties of the superparamagnetic assembly of magnetically interacting MNPs. In the kMC simulation algorithm, there exists a linear relationship between the simulation steps and the real time scale, which holds quite well for a wide range of frequencies~\cite{chantrell2000}. Therefore, the simulated relaxation curves are obtained as a function of time, facilitating the precise evaluation of relaxation time. In the Metropolis Monte Carlo method, on the other hand, the relationship between simulation step and time is not well defined, so the magnetic relaxation curves are obtained as a function of Monte Carlo steps, precluding a quantitative comparison with analytical models~\cite{melenev2012}. The kMC simulation has been used to study the various quantities of interest in MNPs assembly~\cite{anand2019,tan2014,carrey2016,anand2020}. For instance, Tan {\it et al.} studied the spatial dependence of heat dissipation due to hysteresis in an assembly of dipolar interacting MNPs~\cite{tan2014}. We have recently implemented the kMC algorithm to analyze the hysteresis response in a one-dimensional chain of MNPs as a function of dipolar interaction strength and anisotropy axis orientation~\cite{anand2020}. It has also been used to study transport properties~\cite{muller2011,verdes2002}.

The remaining of the paper is organized as follows: In Sec.~II, we present the model and discuss the various energy terms. We also discuss the kMC simulations briefly. The simulation results will be discussed in Sec. III. Finally, we summarize and conclude the work in Sec.~IV.

\section{Model}
\begin{figure}[!htb]
	\centering\includegraphics[scale=0.50]{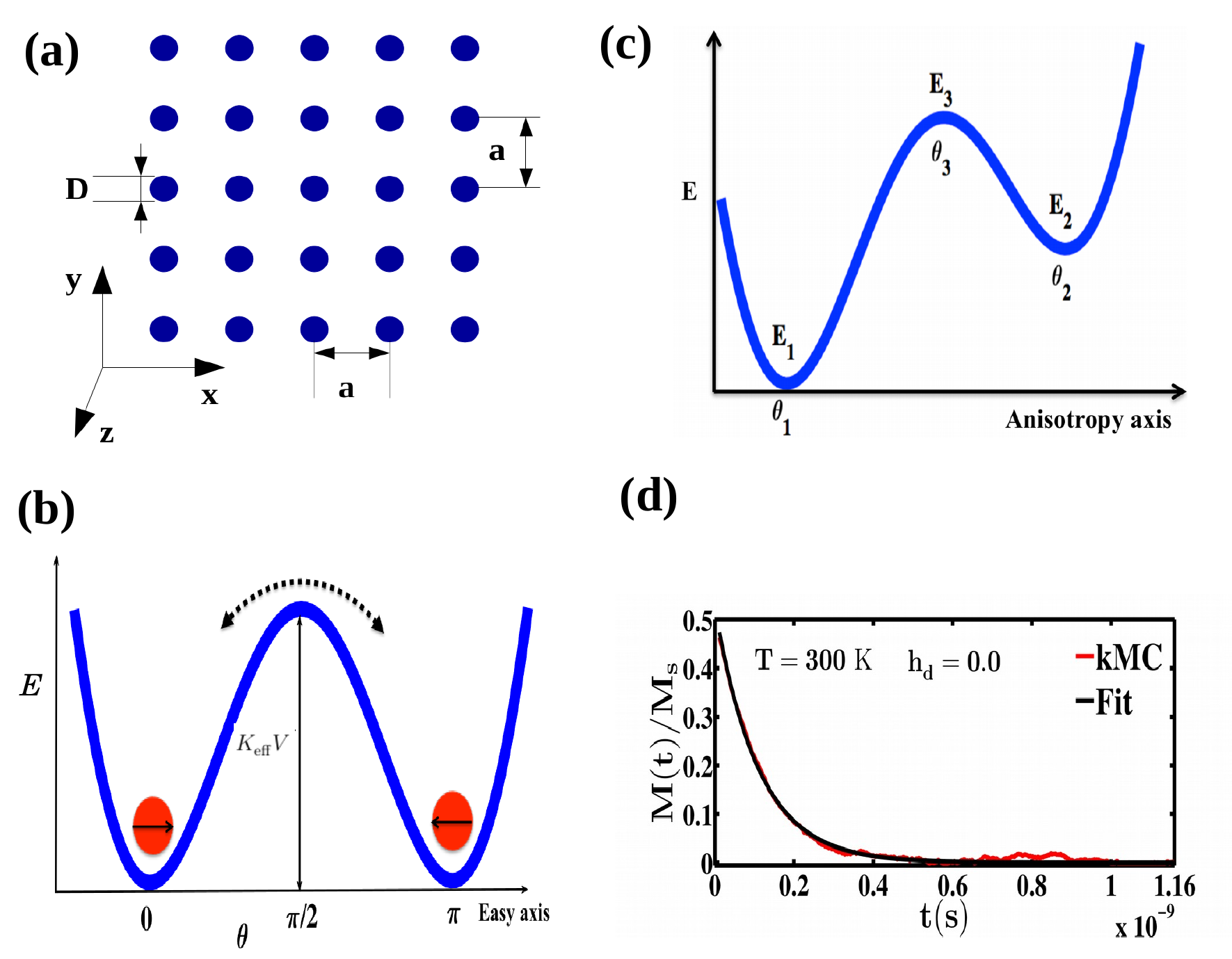}
	\caption{(a) Schematic of the two-dimensional array of magnetic nanoparticles. $a$ is the lattice constant, and $D$ is the particle diameter. (b) Schematic of the energy barrier in the absence of dipolar interaction. There are the two energy minima at $\theta=0$ and $\pi$,  respectively. There is an energy maximum  of strength $K_{\mathrm {eff}} V$ at $\theta=\pi/2$. (c) Schematic of the energy barrier in the presence of dipolar interaction. The modified energy minima are $E^{}_1$ and $E^{}_2$ and the maximum is $E^{}_3$. (d) Simulated magnetization decay $M(t)/M^{}_s$ versus $t$ curve for $h^{}_d=0.0$ at temperature $T=300$ K. It has been fitted with $M(t)=M^{}_s\exp(-t/\tau^{0}_N)$ and shown with the black line. The fitted value comes out to be $\tau^{0}_N=1.161\times10^{-10}\pm 1.10\times10^{-11}$ s. The theoretical value of  $\tau^{0}_N$ is $1.160\times10^{-10}$ s. It shows perfect agreement between the theory and kMC simulation.}
	\label{figure1}
\end{figure}

We consider an assembly of spherically shaped and monodisperse magnetic nanoparticle arranged in a two-dimensional lattice in the $xy$-plane with system dimension $L^{}_x\times L^{}_y$. The particle has a diameter $D$, and the lattice spacing is $a$, as shown in Fig.~\ref{figure1}(a). Each particle has a magnetic moment $\mu=M^{}_sV$, $M^{}_s$ being the saturation magnetization, and $V=\pi D^3/6$ is the volume of the nanoparticle. We have also assumed that each nanoparticle has uniaxial anisotropy constant $K^{}_{\mathrm {eff}}$ is the uniaxial anisotropy constant. The anisotropy or the easy axes are considered to have random orientations to mimic the actual system. The energy of single domain superparamagnetic nanoparticle due to uniaxial anisotropy is given by~\cite{anand2019,dattagupta2012,muscas2018} 
\begin{equation}
E=K^{}_\mathrm {eff} V\sin^2\theta.
\label{barrier}
\end{equation}
Here $\theta$ is the angle between the magnetic moment and the anisotropy axis. It is evident from Eq.~(\ref{barrier}) that the energy function of a single nanoparticle is a symmetric double well. There are two energy minima $E^{0}_1$ and $E^{0}_2$ at $\theta=0$ and $\pi$, respectively. These energy minima are separated by an energy barrier $E^{0}_3=K^{}_\mathrm {eff} V$ at $\theta=\pi/2$ as depicted in Fig.~\ref{figure1}(b). Due to thermal fluctuations, magnetic moment changes its orientation within the particle by overcoming the energy barrier $K^{}_\mathrm {eff} V$. The mean time taken by the magnetic moment to change its direction is known as N\'eel relaxation time $\tau^{0}_{N}$ defined as~\cite{anand2019,carrey2011}
\begin{equation}
\tau^{0}_{N}=\tau^{}_o\exp(K^{}_\mathrm {eff}V/k^{}_BT).
\label{Neel}
\end{equation}
Here $\tau^{}_o=(2\nu_o)^{-1}$, $\nu_o\approx10^{10}$ $s^{-1}$ is the attempt frequency. $k^{}_B$ is the Boltzmann constant and $T$ is the temperature. Eq.~(\ref{Neel}) is applicable for non-interacting MNPs.

In an assembly, MNPs primarily interact due to long-ranged dipolar interaction. We can calculate the dipolar interaction energy $E^{}_{\mathrm {dip}}$ in such a case as~\cite{odenbach2002,usov2017,anand2021hys}
\begin{equation}
\label{dipole}
E^{}_{\mathrm {dip}}=\frac{\mu^{}_o}{4\pi a^3}\sum_{j,\ j\neq i}\left[ \frac{\vec{\mu_{i}}\cdot\vec{\mu_{j}}}{(r^{}_{ij}/a)^3}-\frac{3\left(\vec{\mu_{i}}\cdot\hat{r}_{ij}\right)\left(\vec{\mu_{j}}\cdot\hat{r}_{ij}\right)}{(r^{}_{ij}/a)^3}\right].
\end{equation}
Here $\mu_{o}$ is the permeability of free space; $\vec{\mu}_{i}$ and $\vec{\mu}_{j}$ are the magnetic moment vectors of $i^{th}$ and $j^{th}$ nanoparticle, respectively, and $r^{}_{ij}$ is the center-to-center separation between $\mu_{i}$ and $\mu_{j}$.  $\hat{r}^{}_{ij}$ is the unit vector corresponding to $\vec{r}_{ij}$.
The corresponding dipolar field $\mu^{}_o\vec{H}^{}_{\mathrm {dip}}$ is given by the following expression~\cite{anand2021hys,tan2014}
\begin{equation}
\mu^{}_{o}\vec{H}^{}_{\mathrm {dip}}=\frac{\mu\mu_{o}}{4\pi a^3}\sum_{j,j\neq i}\frac{3(\hat{\mu}^{}_j \cdot \hat{r}_{ij})\hat{r}^{}_{ij}-\hat{\mu^{}_j} }{(r_{ij}/a)^3}.
\label{dipolar1}
\end{equation}
As it is evident from Eq.~(\ref{dipole}) and Eq.~(\ref{dipolar1}) that dipolar interaction varies as $1/r^{3}_{ij}$, we can define the strength of dipolar interaction $h^{}_d=D^{3}/a^3$~\cite{tan2010}. So $h^{}_d= 1.0$ is the largest dipolar interaction
strength, and $h^{}_d=0$ can be termed as the non-interacting case.
Therefore, the total energy of the underlying system is given by~\cite{tan2014,anand2019}
\begin{equation}
E=K^{}_{\mathrm {eff}}V\sum_{i}\sin^2 \theta^{}_i+\frac{\mu^{}_o\mu^2}{4\pi a^3}\sum_{j,\ j\neq i}\left[ \frac{\hat{\mu_{i}}\cdot\hat{\mu_{j}}-{3\left(\hat{\mu_{i}}\cdot\hat{r}_{ij}\right)\left(\hat{\mu_{j}}\cdot\hat{r}_{ij}\right)}}{(r_{ij}/a)^3}\right]
\end{equation}
Here $\theta^{}_i$ is the angle between the anisotropy axis and the $i^{th}$ magnetic moment of the system.

The energy function given by Eq.~(\ref{barrier}) is modified
due to the dipolar interaction. Consequently, the energy barrier seen by the moments is also altered. The single-particle energy function defined by Eq.~(\ref{barrier}) becomes asymmetric, as depicted in Fig.~\ref{figure1}(c). When the dipolar field is greater than the single-particle anisotropy field $H^{}_K=2K^{}_{\mathrm {eff}}/M^{}_s$~\cite{carrey2011}; the energy function has only one minimum. The energy profile displays two minima $E^{}_1$ and $E^{}_2$, and a maxima $E^{}_3$ for each magnetic moment when the dipolar field is less than $H^{}_K$, as shown in Fig.~\ref{figure1}(c). Therefore, the jump rate $\nu^{}_1$ for the magnetic moment to go from $E^{}_1$ to $E^{}_2$ via $E^{}_3$ is given by ~\cite{hanggi1990}
\begin{equation}
\nu^{}_1=\nu^{0}_{1}\exp\bigg(-\frac{E^{}_3-E^{}_1}{k^{}_BT}\bigg)
\end{equation}
Similarly the jump rate $\nu^{}_2$ for the magnetic moment to switch its orientation from $E^{}_2$ to $E^{}_1$ is expressed as~\cite{hanggi1990}
\begin{equation}
\nu^{}_2=\nu^{0}_2\exp\bigg(-\frac{E^{}_3-E^{}_2}{k^{}_BT}\bigg),
\end{equation} 
where $\nu^{0}_{1}=\nu^{0}_{2}=\nu^{}_{o}$. 

The kMC procedure used in the present work is described in greater detail in the work of Anand {\it et al.} and Tan {\it et al.}~\cite{anand2019,tan2014}. Therefore, we do not describe it here to avoid repetition. The kMC simulation has been used to calculate effective N\'eel relaxation time $\tau^{}_N$ for dipolar interacting MNPs by studying magnetization $M(t)$ decay as a function of time $t$. For this, we apply a very large external magnetic field $\mu_oH_{\mathrm{max}}=20$ Tesla along the $y$-direction with respect to the underlying system so that all the magnetic moments point along the applied field direction. The total simulation time is divided into 2000 equal time steps, and we switch off $\mu_oH_{\mathrm{max}}$ at $t=0$ to study the magnetic relaxation  as a function of time. As time passes, magnetic moments relax to a state with zero or an extremely small magnetization. The resulting magnetization-decay curve is then fitted to the form $M(t) = M^{}_s \exp(-t/\tau^{}_N)$ to extract $\tau^{}_N$.

\section{Simulations Results}
We consider spherical  nanoparticles of Fe$_3$O$_4$ arranged in a two-dimensional lattice ($L^{}_x\times L^{}_y$) with $D=8$ nm, $K_{\mathrm {eff}}=13\times10^3$ Jm$^{-3}$, and $M^{}_s=4.77\times10^5$ Am$^{-1}$. The total number of nanoparticles in the assembly is considered as $n=400$. We have considered seven values of system sizes viz. $L_x\times L_y=20\times20$, $16\times25$, $10\times40$, $8\times50$, $4\times100$, $2\times200$ and $1\times400$. The corresponding aspect ratio $A^{}_r(=L^{}_y/L^{}_x)$ of the underlying system is $1.0$, 1.56, 4.0, 6.25, 25, 100 and 400, respectively. The dipolar interaction strength $h^{}_d$ is varied from 0 to 1.0. The temperature $T$ is changed between 100 and 400 K. The anisotropy axes are assumed to have random orientations. All the numerical data obtained have been averaged over several independent runs to obtain good statistical averaging. 

To validate our kMC method, we first study the magnetic relaxation in the absence of dipolar interaction at room temperature. In Fig.~\ref{figure1}(d), we plot the simulated magnetization decay $M(t)/M^{}_s$ versus $t$ curve of a square arrangement of MNPs ($L^{}_x\times L^{}_y=20\times20$) with $h^{}_d=0.0$ and $T=300$ K. The functional form of the curve is a perfectly exponential decay. We fit it with $M(t)/M^{}_s=\exp(-t/\tau^{0}_{N})$ which yields $\tau^{0}_N=1.161\times10^{-10}\pm 1.10\times10^{-11}$ s. The theoretical value calculated using Eq.~(\ref{Neel}) comes out to be $\tau^{0}_N=1.160\times10^{-10}$ s, which is in perfect agreement with the simulated one. In the absence of magnetic interaction, the magnetization decay curve is also found to be independent of the aspect ratio $A^{}_r$ as expected. Therefore, we have not shown the corresponding curves to avoid duplication.

\begin{figure}[!htb]
	\centering\includegraphics[scale=0.40]{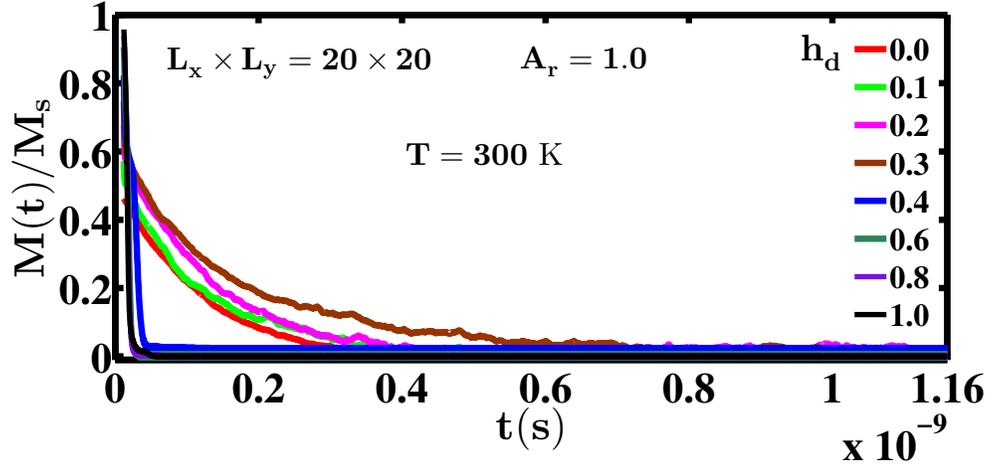}
	\caption{Magnetization decay curve as a function of dipolar interaction strength $h^{}_d$ for square assembly of MNPs at $T=300$ K. There is an increase in magnetization relaxation with an increase in $h^{}_d$ provided dipolar interaction strength is appreciable ($h^{}_d>0.3$). It can be attributed to an enhancement in antiferromagnetic coupling with an increase in interaction strength.}
	\label{figure2}
\end{figure}

Next, we study the effect of dipolar interaction on the magnetic relaxation with the square arrangement of MNPs at $T=300$ K. In Fig.~(\ref{figure2}), we plot $M(t)/M^{}_s$ versus $t$ curve with $L^{}_x \times L^{}_y=20\times20$ ($A^{}_r=1.0$) for eight representative values of $h^{}_d=$ 0.0, 0.1, 0.2, 0.3, 0.4, 0.6, 0.8 and 1.0. There is a smooth decay of magnetization in the presence of weak dipolar interaction $h^{}_d\leq0.3$. The functional form of the magnetization decay is perfectly exponential decaying in this case also. Interestingly, there is a fastening in magnetization relaxation with an increase in dipolar interaction strength for strongly interacting MNPs ($h^{}_d>0.3$). It could be explained using the fact that the dipolar interaction induces antiferromagnetic coupling between the magnetic moments in a square lattice. The strength of antiferromagnetic coupling increases with an increase in $h^{}_d$, which results in the fastening of magnetization relaxation. The dominance of antiferromagnetic coupling is in perfect agreement with our recent work, where we have also found a characteristic hysteresis curve of antiferromagnetic coupling predominance in a square lattice~\cite{anand2021hys}. De'Bell {\it et al.} also observed dominance of antiferromagnetic interaction with the Heisenberg spins arranged in square array~\cite{de1997}. The functional form of magnetization decay for weakly interacting MNPs is also in qualitative agreement with the work of Figueiredo {\it et al.}~\cite{figueiredo2007}.

\begin{figure}[!htb]
\centering\includegraphics[scale=0.40]{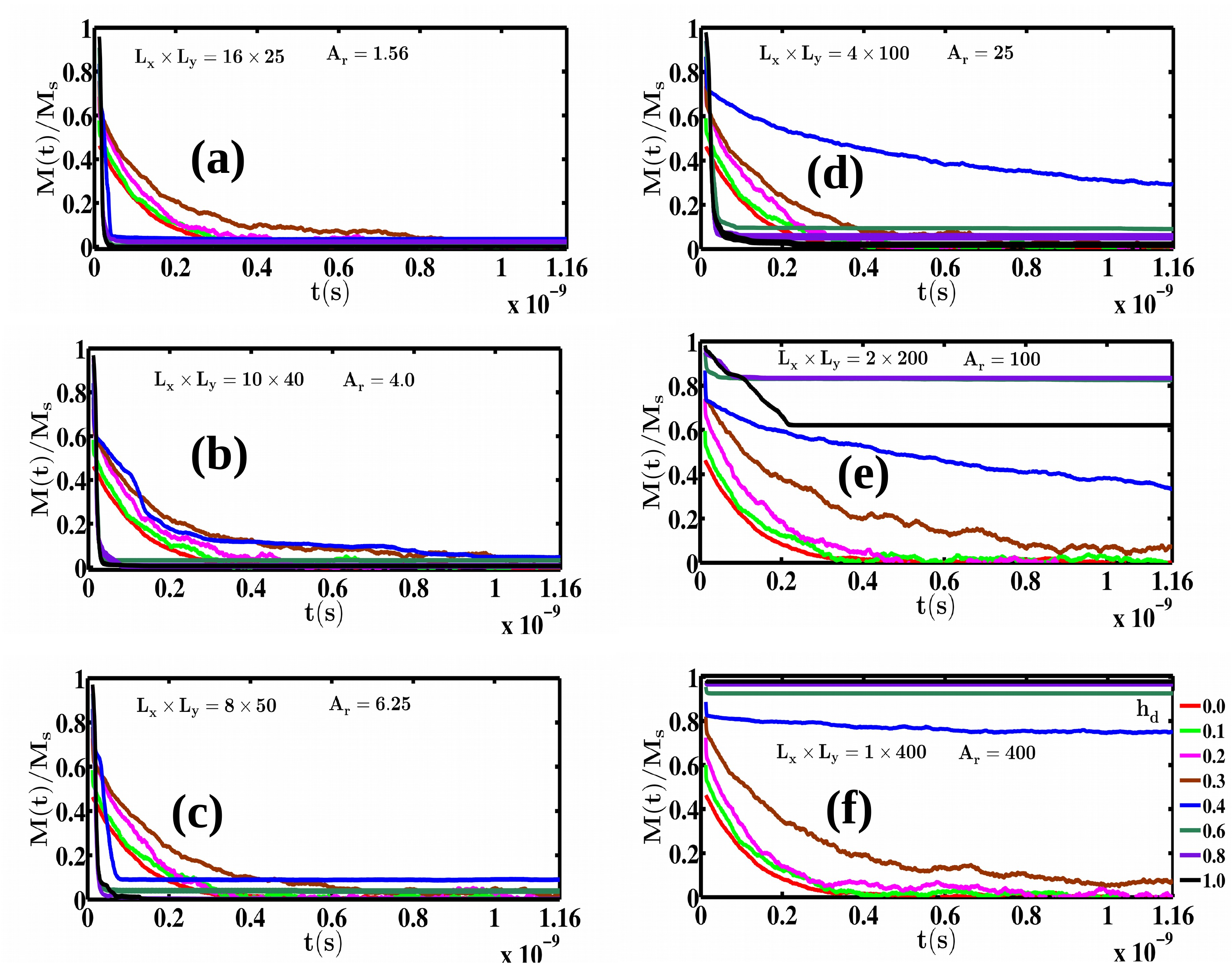}
\caption{Magnetization decay $M(t)/M^{}_s$ versus $t$ curves as a function of dipolar interaction strength $h^{}_d$ for various values of aspect ratio $A^{}_r$ at $T=300$ K. It is evident that there is an increase in magnetic relaxation with $h^{}_d$ even with the rectangular arrangement of MNPs ($A^{}_r>1$), which can be due to an increased in antiferromagnetic coupling. Interestingly, there is a slowing down in magnetization decay for huge aspect ratio $A^{}_r$ because of enhancement in ferromagnetic interaction.} 
\label{figure3}
\end{figure}

We then study the effect of the shape of the system, i.e., aspect ratio $A^{}_r$ on the magnetic relaxation. In Fig.~(\ref{figure3}), we plot the magnetization decay $M(t)/M^{}_s$ vs. $t$ curve for six representative values of $A^{}_r=1.56$, 4.0, 6.25, 25, 100 and 400. The dipolar interaction strength $h^{}_d$ has been varied between 0.0 to 1.0. Irrespective of $A^{}_r$, the functional form of the magnetization decay is perfectly exponentially decaying for weakly interacting MNPs ($h^{}_d\le0.3$). Remarkably, there is a fastening of magnetization relaxation above certain dipolar interaction strength $h^{\star}_d$, which also increases with $A^{}_r$. It is seen that there is a slowing down of magnetization relaxation below $h^{\star}_d$. On the other hand, magnetization relaxes rapidly above this threshold value of $h^{\star}_d$, which can be due to enhancement in antiferromagnetic coupling. In the case of enormous $A^{}_r$, the system behaves as highly anisotropic. Consequently, the dipolar interaction promotes ferromagnetic coupling, which slows down the magnetization reversal (slow decay of magnetization) with an increase in $h^{}_d$. The slowing down of magnetic relaxation with  dipolar interaction strength in the case of $A^{}_r=400$ is in perfect agreement with the work of Iglesias {\it et al.}~\cite{iglesias2004}. These observations indicate that one can tune the nature of dipolar interaction from ferromagnetic to antiferromagnetic by just varying  the aspect ratio of the system. Consequently, the relaxation time can be manipulated in a more controlled way, which is an essential aspect of data storage applications.

\begin{figure}[!htb]
\centering\includegraphics[scale=0.40]{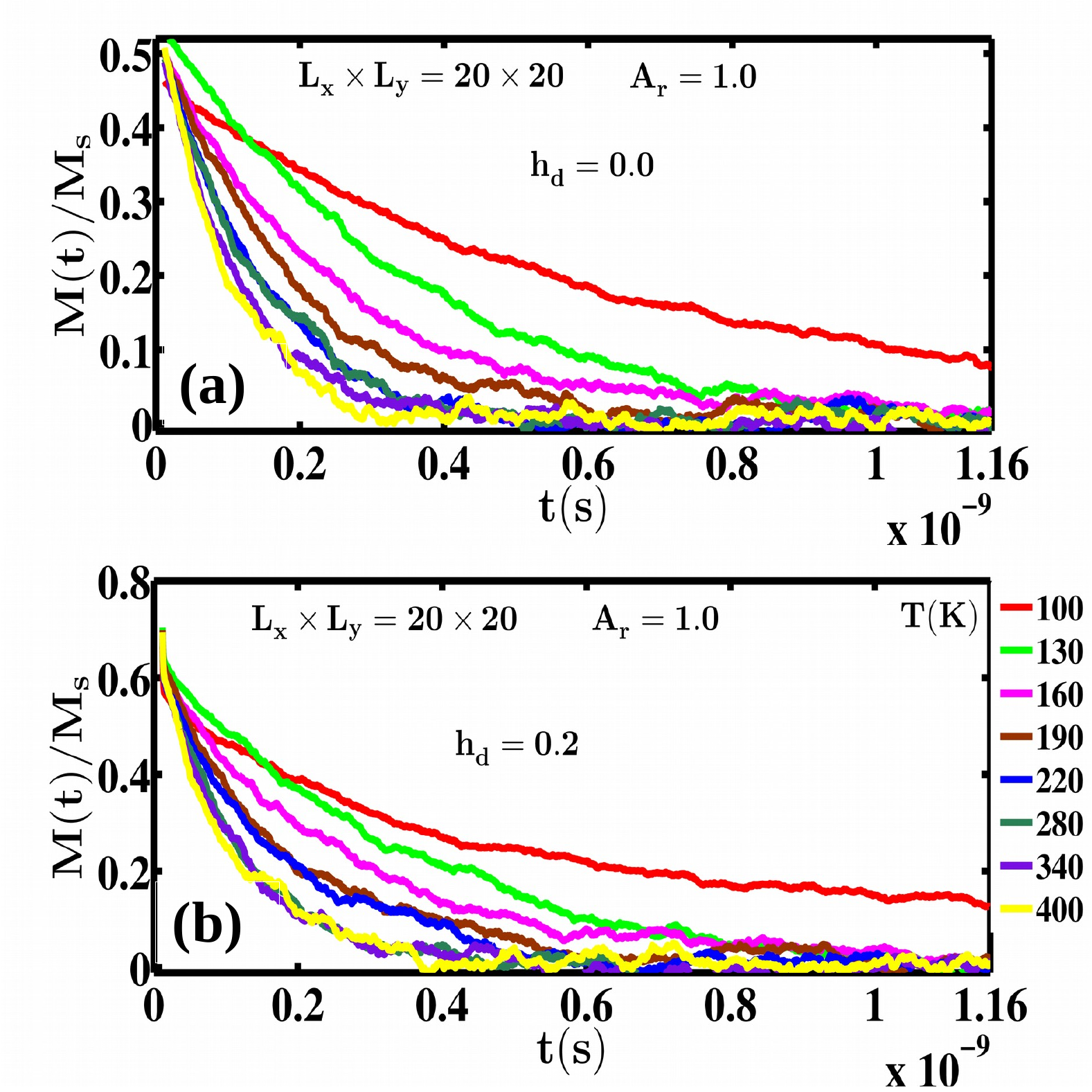}
\caption{Magnetization decay $M(t)/M^{}_s$ versus $t$ curve as a function of temperature $T$ for weakly interacting MNPs. We have considered two value of dipolar interaction strength $h^{}_d=0.0[(a)]$ and 0.2[(b)]. In the case of non-interacting MNPs ($h^{}_d=0.0$) and a given temperature, magnetization relaxes faster as compared to h$_d=0.2$. Magnetization-decay does not depend on aspect ratio $A^{}_r$ as expected (curves not shown).}
\label{figure4}
\end{figure}

It is equally important to understand the effect of thermal fluctuations on magnetic relaxation. In Fig.~(\ref{figure4}), we plot the magnetization decay curve as a function of temperature $T$ for weak dipolar interaction. We have considered the square arrangement of MNPs ($L^{}_x\times L^{}_y=20\times20$, $A^{}_r=1.0$) with two values of dipolar interaction strength $h^{}_d=0.0$ and 0.2. There is a smooth decay of magnetization, and the functional form of the magnetic relaxation curve is also perfectly exponential irrespective of $T$ and $h^{}_d$. There is a slow decay of magnetization in the absence of dipolar interaction and smaller values of $T$. There is a fastening in magnetization relaxation with an increase in temperature. It is due to the enhancement in thermal fluctuations. The rate of magnetization decay becomes slower for comparably large dipolar interaction strength ($h^{}_d=0.2$). In the presence of weak interaction, there is always an increase in magnetization decay with increased thermal fluctuations. The functional form of magnetization decay is also found to be independent of aspect ratio $A^{}_r$ of the system  for weak dipolar interaction, as expected (curves not shown).

\begin{figure}[!htb]
\centering\includegraphics[scale=0.40]{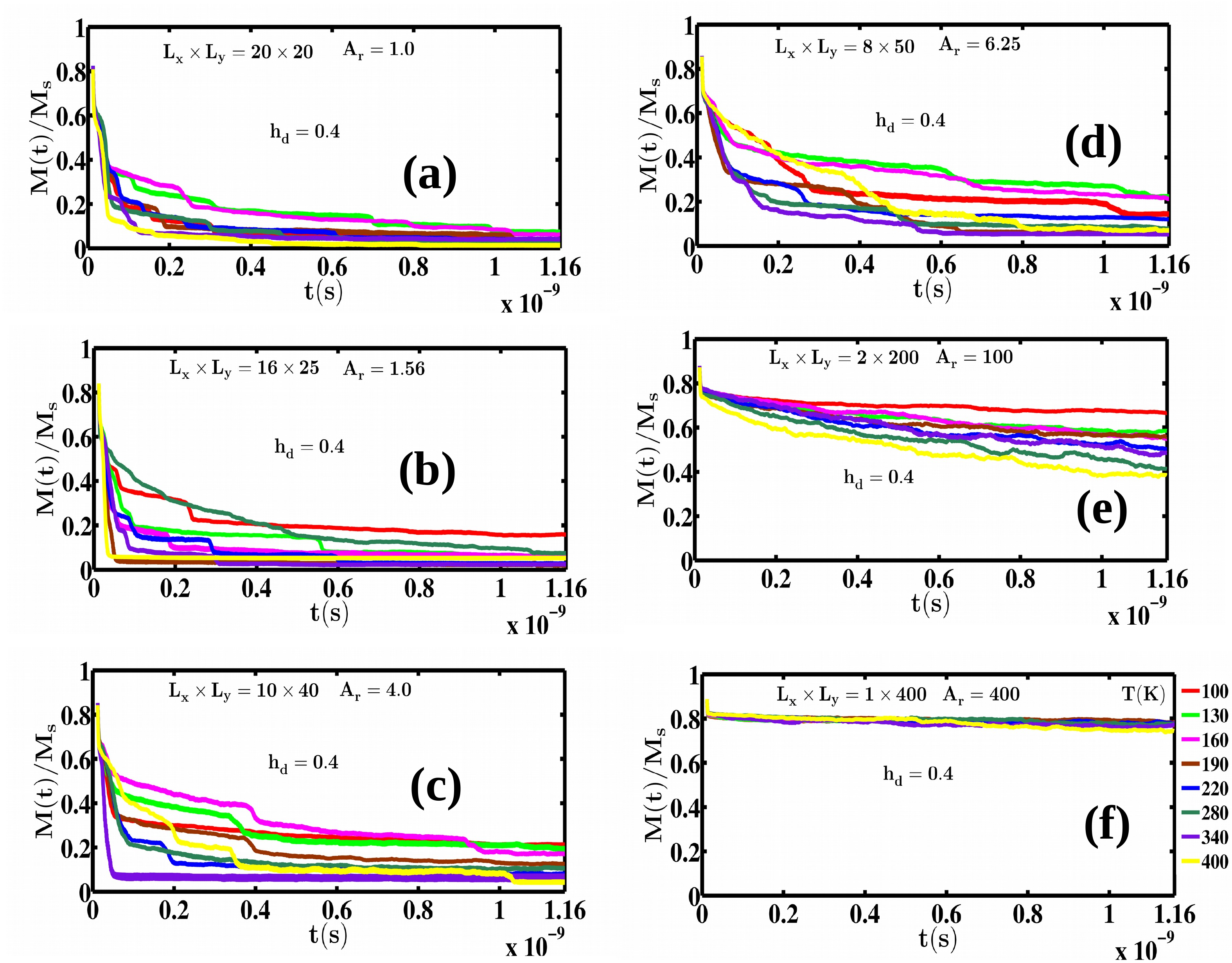}
\caption{Magnetization-decay curves as a function of temperature $T$ for $h^{}_d=0.4$. We have considered six values of aspect ratio $A^{}_r=1.0$[(a)], 1.56[(b)], 4.0[(c)], 6.25[(d)], 100[(e)] and 400[(f)]. For $A^{}_r<100$, magnetization decays rapidly as compared with the non-interacting case. It is because the nature of dipolar interaction changes from antiferromagnetic to ferromagnetic for exceedingly large $A^{}_r$. There is also an increase in magnetization relaxation with temperature for $A_r<100$. While for a huge aspect ratio, there is a weak dependence of relaxation on thermal fluctuations.}
\label{figure5}
\end{figure}

The magnetic relaxation should depend on the aspect ratio for strongly  interacting MNPs. Therefore, we now study the magnetization decay as a function of temperature $T$ with various values of $A^{}_r$ in the presence of large dipolar interaction strength. In Fig.~(\ref{figure5}), we plot the magnetization decay $M(t)/M^{}_s$ vs. $t$ with $h^{}_d=0.4$ and six values of $A^{}_r=$1.0, 1.56,4.0,6.25, 100 and 400. The temperature has been varied between 100 and 400 K. It is seen that magnetization decay is faster as compared to the weakly interacting MNPs (Fig.~(\ref{figure4})) for a given temperature and $A^{}_r<100$. Magnetization also decays faster than an exponential function. It means that the large dipolar interaction strength promotes antiferromagnetic interaction. With an increase in thermal fluctuations, there is fastening in magnetization relaxation. Interestingly, magnetization decays very slowly with $A^{}_r>100$ and also has a weak temperature dependence. The dipolar interaction induces ferromagnetic coupling between the magnetic moments for the highly anisotropy system, i.e. for huge $A^{}_r$. As a consequence, magnetization ceases to relax for large dipolar interaction strength.

\begin{figure}[!htb]
\centering\includegraphics[scale=0.40]{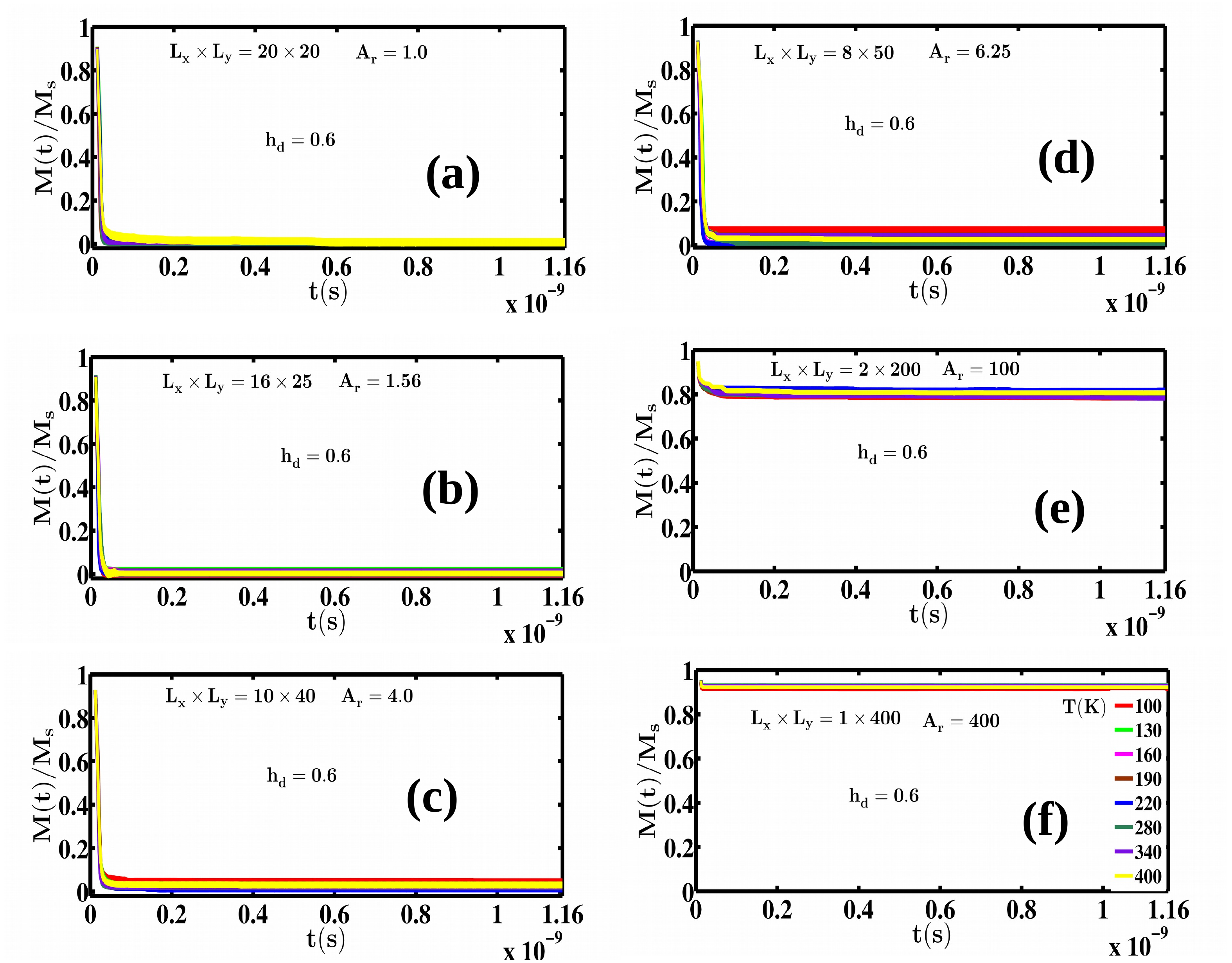}
\caption{Magnetization decay $M(t)/M^{}_s$ versus $t$ curve as a function of temperature $T$ for strongly interacting MNPs ($h^{}_d=0.6$). We have considered six values of aspect ratio $A^{}_r=1.0$[(a)], 1.56[(b)], 4.0[(c)], 6.25[(d)], 100[(e)] and 400[(f)]. In the case of $A^{}_r<100$, magnetization relaxes very rapidly as compared to non-interacting case. It is because the dipolar interaction promotes antiferromagnetic coupling between the MNPs in such a case. On the other hand, MNPs ceases to relax with a huge aspect ratio ($A^{}_r>100$) due to the enhanced ferromagnetic interaction. There is a weak dependence of relaxation on thermal fluctuations.}
\label{figure6}
\end{figure}

\begin{figure}[!htb]
\centering\includegraphics[scale=0.40]{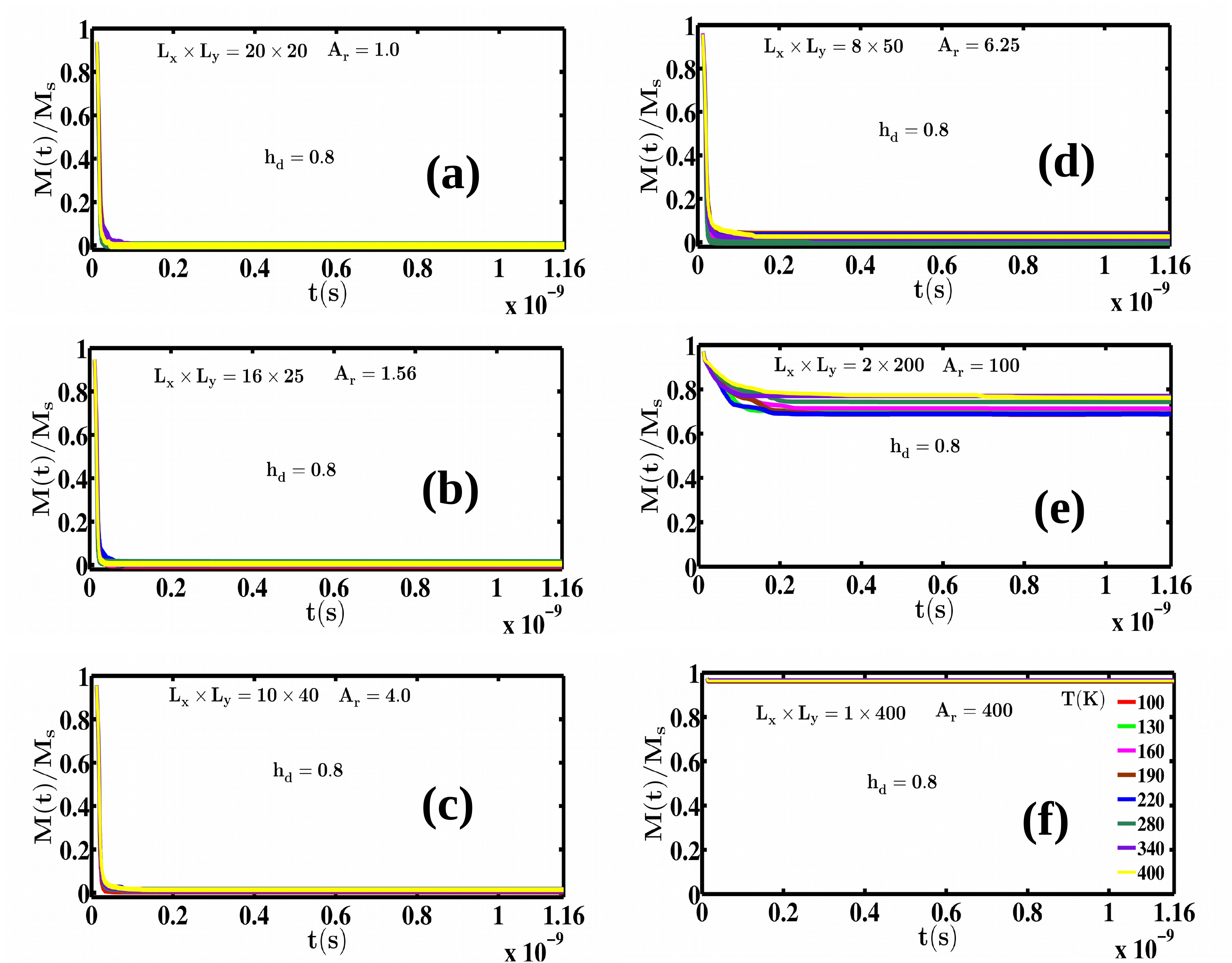}
\caption{Magnetization decay curves as a function of temperature $T$ for $h^{}_d=0.8$. We have considered six values of aspect ratio $A^{}_r=1.0$[(a)], 1.56[(b)], 4.0[(c)], 6.25[(d)], 100[(e)] and 400[(f)]. There is a rapid fall in magnetization as a function of time for $A^{}_r<100$ because of antiferromagnetic interaction indued by dipolar interaction. The dipolar interaction promotes ferromagnetic coupling between the MNPs with $A^{}_r>100$, resulting in the slow decay of magnetization. Temperature does not affect the relaxation characteristics because of the large dipolar interaction strength.}
\label{figure7}
\end{figure}

Next, we study the variation of magnetization decay $M(t)/M^{}_s$ versus $t$ curve as a function of temperature for various values of $A^{}_r$ and very large dipolar interaction strength $h^{}_d=0.6$ and 0.8 in Fig.~(\ref{figure6}) and Fig.~(\ref{figure7}), respectively. All the other parameters are the same as that of Fig.~(\ref{figure5}). For $A^{}_r<100$, the magnetization decays very fast as compared to the weakly interacting case. It is due to the fact the antiferromagnetic coupling is enhanced because of dipolar interaction in such cases. Therefore, magnetization tends to change its orientation very rapidly, resulting in rapid decay of magnetization. On the other hand, the dipolar interaction of equal strength induces ferromagnetic coupling when the aspect ratio is enormous ($A^{}_r>100$). Consequently, magnetization relaxes very slowly or does not relax at all for sizeable dipolar interaction strength. We observe weak dependence of magnetization relaxation on thermal fluctuations provided interaction strength is immense. The rapid fall of magnetization as a function of time is in perfect qualitative agreement with the work of Volkov {\it et al.}~\cite{volkov2008}. We could not compare the entire range of dipolar interaction strength and temperature as they have concentrated on a single value of interaction strength temperature. Therefore our results can be used as a benchmark in this context.
We believe that these results will help the experimentalist in choosing suitable values of system size, interaction strength and other parameters of interest for desired relaxation characteristics for better applications of these systems in data storage and other related applications.

\begin{figure}[!htb]
	\centering\includegraphics[scale=0.40]{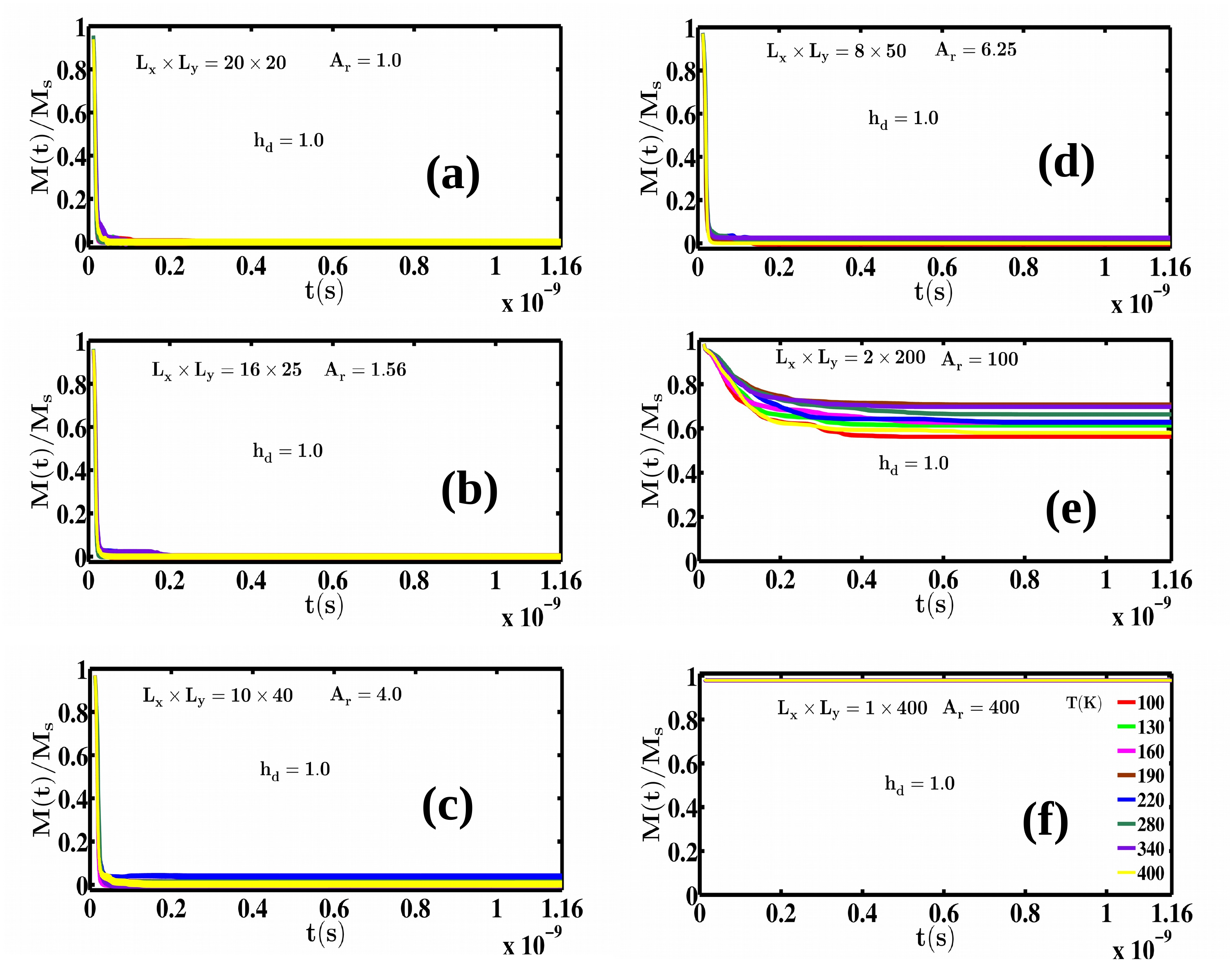}
	\caption{Magnetization-decay $M(t)/M^{}_s$ versus $t$ curve as a function of temperature $T$ for the strongest dipolar interacting MNPs ($h^{}_d=1.0$). We have considered six values of aspect ratio $A^{}_r=1.0$[(a)], 1.56[(b)], 4.0[(c)], 6.25[(d)], 100[(e)] and 400[(f)]. Magnetization relaxes very rapidly with time for $A^{}_r<100$.  On the other hand, magnetization relaxes slowly or does relax at all for $A^{}_r>100$ because of large ferromagnetic interaction. Thermal fluctuations does not affect the relaxation as dipolar interaction strength is the largest.}
	\label{figure8}
\end{figure}

\begin{figure}[!htb]
	\centering\includegraphics[scale=0.45]{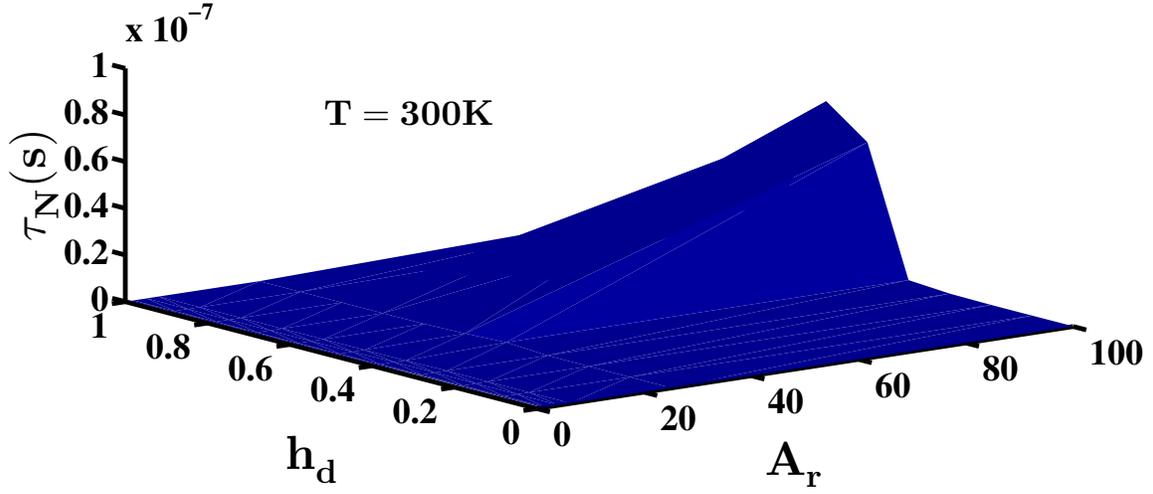}
	\caption{Variation of N\'eel relaxation time $\tau^{}_N$ as a function of dipolar interaction strength $h^{}_d$ and aspect ratio $A^{}_r$ at T=300 K. $\tau^{}_N$ decreases above a particular value of dipolar interaction strength $h^{\star}_d$. The latter also increases with $A^{}_r$. In the case of enormous $A^{}_r$, $\tau_N$ always increases with $h_d$ because of an enhancement in ferromagnetic coupling.}
\label{figure9}
\end{figure}

In Fig.~(\ref{figure8}), we plot the magnetization decay curve as a function of temperature for the strongest dipolar interacting MNPs ($h^{}_d=1.0$). All other parameters are the same as that of Fig.~(\ref{figure7}). As the antiferromagnetic coupling is the maximum, the decay rate of magnetization is the fastest for $A^{}_r<100$. Remarkably, the ferromagnetic interaction is also the largest for highly anisotropic system ($A^{}_r>100$). Therefore, magnetization ceases to relax, which results in slowing down of magnetization relaxation for the highly anisotropic system such as linear chain of MNPs. The temperature has a negligible effect on the magnetic relaxation as the dipolar interaction strength  is the strongest. To study the effect of dipolar interaction strength and aspect ratio on the magnetic relaxation quantitatively, we now plot the variation of simulated N\'eel relaxation time $\tau^{}_N$ as a function of $h^{}_d$ and $A^{}_r$ at $T=300$ K in Fig.~(\ref{figure9}). $\tau^{}_N$ is extracted from the simulated magnetization decay $M(t)/M^{}_s$ versus $t$ curve by fitting it with $M(t)/M^{}_s=\exp(-t/\tau^{}_N)$. There is a decrease in $\tau^{}_N$ above a particular value of dipolar interaction strength $h^{\star}_d$. There is also an increase in $h^{\star}_d$ with $A^{}_r$. It is quite evident that $\tau^{}_N$ increases with dipolar interaction strength below $h^{\star}_d$. On the other hand, $\tau^{}_N$ decreases very fast above this threshold value of $h^{\star}_d$, which can be due to enhancement in antiferromagnetic coupling. In the case of extremely large $A^{}_r$, the system behaves as highly anisotropic. Therefore, $\tau^{}_N$ always increases with $h^{}_d$ because of an enhancement in ferromagnetic coupling in these cases.

\begin{figure}[!htb]
	\centering\includegraphics[scale=0.40]{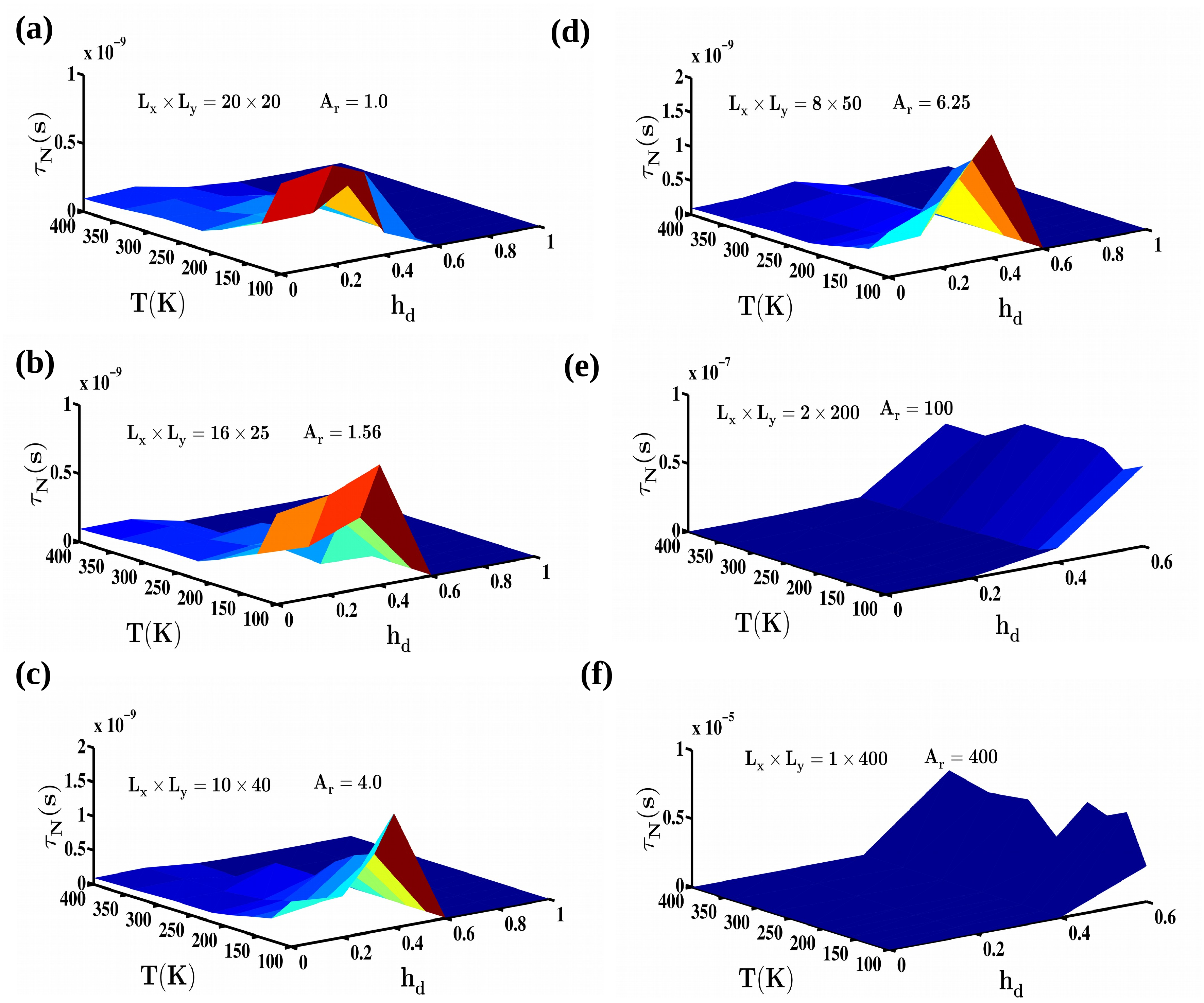}
	\caption{Variation of N\'eel relaxation time $\tau^{}_N$ as a function of dipolar interaction strength $h^{}_d$ and $T$ for six values of $A^{}_r=1.0$[(a)], 1.56[(b)], 4.0[(c)], 6.25[(d)], 100[(e)], and 400[(f)]. $\tau^{}_N$ decreases with $h_d$ for appreciable dipolar interaction strength ($h^{}_d > 0.3$) and $A_r<100$. Remarkably, $\tau_N$ increases with $h_d$ for the highly anisotropic system
	($A_r >100$). It is because the dipolar interaction induces ferromagnetic coupling in such cases. There is a rapid fall in $\tau^{}_N$ with $T$ because of enhancement in thermal fluctuations.} 
\label{figure10}
\end{figure}

Finally, we study the variation of $\tau^{}_N$ as a function of temperature and dipolar interaction strength for various values of $A^{}_r$. In Fig.~(\ref{figure10}), we plot the variation of $\tau^{}_N$ as a function of $T$ and $h^{}_d$ for six representative values of $A^{}_r=1.0$, 1.56, 4.0, 6.25, 100 and 400. It is evident that $\tau^{}_N$ decrease with $h^{}_d$ for considerable dipolar interaction strength ($h^{}_d>0.3$) and $A^{}_r<100$. It is because the dipolar interaction promotes antiferromagnetic coupling for appreciable dipolar interaction strength and $A^{}_r<100$. As a consequence, there is a decrease in $\tau^{}_N$ with $h^{}_d$. Remarkably, $\tau^{}_N$ increases with $h^{}_d$ for the highly anisotropic system ($A^{}_r>100$). It is due to the fact that dipolar interaction induces ferromagnetic coupling in such cases. In the case of weakly interacting MNPs, $\tau^{}_N$ decreases rapidly with $T$ because of enhancement in thermal fluctuations. While for strongly interacting MNPs, there is weak dependence of $\tau^{}_N$ on temperature.

\section{Summary and Conclusion}
Now, we summarize and discuss the main results presented in this  work. We have performed extensive kinetic Monte Carlo simulations to probe the magnetic relaxation characteristics in a two-dimensional array of spherical and monodisperse nanoparticle as a function of dipolar interaction strength, the aspect ratio of the system and temperature. In the case of non-interacting and weakly interacting 
MNPs, the functional form of the magnetization-decay curve is a perfect exponential. The extracted N\'eel relaxation time from the simulated curve is also in excellent agreement with the value calculated using the N\'eel-Brown model. Interestingly, there exhibits fastening of magnetic relaxation with an increase in dipolar interaction strength in a square arrangement of MNPs, provided the interaction strength is significant. The same is true even with the large aspect ratio of the system. It can be explained using the fact that dipolar interaction promotes antiferromagnetic coupling between the moments in the square arrangement of MNPs~\cite{macisaac1996}. Consequently, magnetization tends to reverse its orientation very rapidly with an increase in dipolar interaction strength. Therefore, there is fast magnetization relaxation in such cases. Remarkably,  there is a slowing down of magnetization relaxation as interaction strength increases in a highly anisotropic system, i.e., when the aspect ratio is enormous. The dipolar interaction of enough strength induces ferromagnetic interaction between the MNPs in the highly anisotropic system such as a linear chain. Consequently, magnetization ceases to relax in such a situation, resulting in extremely slow magnetization reversal. The thermal fluctuation is also found to affect magnetic relaxation drastically. There is a fastening of magnetization relaxation with temperature in the weak dipolar limit. It is due to the fact that as the temperature is increased, thermal fluctuation increases. Therefore, magnetic moments tend to cross the energy barrier more frequently, which leads to the fastening of magnetic relaxation. 

The effect of dipolar interaction and aspect ratio on the magnetic relaxation is also clearly manifested in the variation of N\'eel relaxation time $\tau^{}_N$. In the presence of strong dipolar interaction and square arrangement of MNPs, $\tau^{}_N$ decreases rapidly with $h^{}_d$ for a given temperature. A similar observation is made for the relatively large aspect ratio $A^{}_r$. It is because the dipolar interaction induces antiferromagnetic coupling in these cases. Remarkably, the dipolar interaction of equal strength promotes ferromagnetic coupling in the highly anisotropic system, such as a linear arrangement of nanoparticles ($A^{}_r$ is enormous). Consequently, $\tau^{}_N$ gets enhanced with $h^{}_d$. There is a rapid increase in the value of $\tau^{}_N$ with an increase in thermal fluctuations for weakly interacting MNPs. On the other hand, $\tau^{}_N$ has a fragile dependence on temperature provided dipolar interaction is strong.

In conclusion, we have studied thermal and dipolar interaction effects on the magnetic relaxation in the two-dimensional arrangement of MNPs with randomly oriented anisotropy axes using kinetic Monte Carlo simulation. Our results suggest that magnetization relaxation can be tuned by just varying the dipolar interaction and shape of the system. However, our results have been obtained for monodisperse magnetic nanoparticles without any position disorder. We believe that similar conclusions can be drawn for the system with the disorder also as long as the dipolar interaction dictates their magnetic behaviour. We also believe that the concepts presented in this work are incredibly relevant for the efficient use of magnetic nanoparticles in data storage and other related applications. These observations could also help the physicist optimize various parameters of interest, such as frequency of the external magnetic field, dipolar interaction strength and shape of the system in magnetic hyperthermia applications.


\bibliographystyle{h-physrev}
\bibliography{ref}

\newpage

\newpage

\newpage

\newpage

\newpage

\newpage

\newpage
\newpage
\newpage
\newpage
\end{document}